\documentstyle[12pt,graphicx,color,subfigure,cite]{article}
\newcommand{\comment}[1]{}
\newcommand{\newc}{\newcommand}

\def\issue(#1,#2,#3){{\bf #1}, #2 (#3)}

\def\PREP(#1,#2,#3){Phys.\ Rep. \issue(#1,#2,#3)}
\def\EPJC(#1,#2,#3){Eur.\ Phys.\ J.\ C \issue(#1,#2,#3)}

\def\thetab0{\theta_{B_0}}

\def\r2{\sqrt 2}
\def\beq{\begin{equation}}
\def\eeq{\end{equation}}
\def\beqn{\begin{eqnarray}}
\def\eeqn{\end{eqnarray}}
\def\sinW2{\sin^2\theta_W}

\def\mz2{M_{z}^2}
\def\c2b{\cos 2\beta}

\def\mz{M_Z}

\def\ptm{E{\!\!\!/}_T}

\def\sec2w{sec^2\theta_W}

\def\gmin2{(g-2)_\mu}

\catcode`\@=11 

\def\lsim{\mathrel{\mathpalette\@versim<}}
\def\gsim{\mathrel{\mathpalette\@versim>}}
\def\@versim#1#2{\vcenter{\offinterlineskip
    \ialign{$\m@th#1\hfil##\hfil$\crcr#2\crcr\sim\crcr } }}

\def\te{\tilde e}

\def\tu{\tilde u}

\def\tb{\tilde b}

\def\tst{\tilde t}

\def\tg{\tilde g}

\newc{\wt}{\widetilde}
\newc{\ra}{\rightarrow}
\newc{\s}{\smallskip}
\newc{\nn}{\noindent}
\newc{\non}{\nonumber}

\def \chonep{{\wt\chi_1}^{+}}
\def \chonem{{\wt\chi_1^-}}
\def \chonep2{{\wt\chi_2^+}}
\def \chonem2{{\wt\chi_2^-}}

\def \lstop{\wt{t}_{1}}

\def \lspone{\wt\chi_1^0}
\def \mlspone{m_{\lspone}}
\def \lsptwo{\wt\chi_2^0}
\def \mlsptwo{m_{\lsptwo}}
\def \lspthree{\wt\chi_3^0}
\def \mlspthree{m_{\lspthree}}
\def \lspfour{\wt\chi_4^0}
\def \mlspfour{m_{\lspfour}}

\def \lsptwo{\wt\chi_2^0}
\def \lspone{\wt\chi_1^0}
\def \chonem {{\wt\chi_1^\pm}}
\def \chargino2 {{\wt\chi_2^\pm}}
\def \lstop{\wt{t}_{1}}
\def \ch2m {{\wt\chi_2^-}}
\def \lspone{\wt\chi_1^0}

\def \chonep {{\wt\chi_1^+}}

\def\mygraph#1#2{ \subfigure[]{
   \label{#1}
   \hspace*{-0.6in}
   \begin{minipage}[b]{0.5\textwidth}
   \centering
   \hspace*{4ex}
   \includegraphics[width=\textwidth]{#2}
   \vspace*{-4ex}
   \end{minipage}}
   \vspace*{-1ex}}
\begin{document}
\begin{flushleft}
\end{flushleft}
\begin{center}
{\large \bf Probing the light Higgs pole resonance annihilation of dark matter in the light of XENON100 and CDMS-II observations. \\}
\vglue 0.5cm
Utpal Chattopadhyay$^{a}$, Debottam Das$^{b,a}$, Dilip Kumar Ghosh$^{a}$ 
and Manas Maity$^{c}$ \\
{$^a$ \em Department of Theoretical Physics, Indian Association 
for the Cultivation of Science,\\  
2A \& B Raja S.C. Mullick Road, Jadavpur, 
Kolkata 700 032, India}\\
{$^b$ \em Laboratoire de Physique Th\'eorique, UMR 8627,
Universit\'e de Paris-Sud 11\& CNRS,\\ B\^atiment 210, 91405 Orsay Cedex,
France}\\
{$^c$ \em Department of Physics, Visva-Bharati, 
Santiniketan 731235, India}
\end{center}


\begin{abstract}
We consider the prospect of lightest neutralino (${\tilde \chi}_1^0$) 
as a dark matter candidate in the light of recent interesting 
observations from the XENON100 and CDMS-II experiments 
in minimal supergravity framework with large $\tan\beta$ 
and nonvanishing $A_0$. Within the WMAP satisfied zone, 
there is a large direct detection reach of lighter ${\tilde \chi}_1^0$ in the 
lighter Higgs boson mediated resonance annihilation domain of the 
above scenario. It is seen that the heavier Higgs boson 
plays a dominating role in 
the ${\tilde \chi}_1^0-p$ cross section in the associated zone of 
parameter space in spite of having a larger mass. 
Possible LHC signatures are discussed.

 \end{abstract}
\noindent
\section{Introduction}
Low energy supersymmetry (SUSY)\cite{SUSY} has several 
features that make it very 
promising as a class of models for beyond the 
standard model (SM) of particle physics\cite{standardmodel}. 
The minimal supergravity (mSUGRA)\cite{msugra} 
is a well-studied model that greatly 
reduces the number of parameters of the minimal supersymmetric standard model 
(MSSM)\cite{SUSY,KaneKingRev} through a few simple requirements. 
The universal input parameters given at the unification scale are: 
(i) the gaugino mass parameter $m_{1/2}$,
(ii) the scalar mass parameter $m_0$, and (iii) the trilinear
SUSY breaking parameter $A_0$. The model that uses radiative 
breaking of electroweak symmetry requires another parameter 
$\tan\beta$ which is the ratio of Higgs vacuum expectation
values. Additionally the sign of the Higgsino mixing parameter $\mu$ is  
an input. With R-parity\cite{KaneKingRev} assumed to be preserved, the model has an 
attractive dark matter\cite{silkphysrep,DMreview} 
candidate, namely the lightest 
neutralino ${\tilde \chi}_1^0$ which is the lightest supersymmetric particle 
(LSP) for most of the parameter space. 
In a scenario where  
${\tilde \chi}_1^0$ is the LSP, the parameter space with other 
sparticles lighter than ${\tilde \chi}_1^0$ are ignored.  
The model is typically associated with 
a Bino-dominated LSP that leads to over-abundance of dark matter. 
Reduction of the relic density to satisfy the 
WMAP data can be achieved if i) there is 
coannihilation of LSP with another sfermion (usually stau ${\tilde \tau}_1$ 
or rarely ${\tilde t}_1$ ) that has mass 
close to the mass of LSP\cite{coannistau}, ii) there is appropriate mixture of 
Bino and Higgsinos in the composition of the LSP 
so that there may be coannihilating charginos 
in the LSP-${\tilde \chi}_1^\pm$ annihilation\cite{Edsjo:1997bg,mizuta},  
the so called 
focus point\cite{focus}/hyperbolic branch\cite{hyper} region,     
 iii) the LSP is 
sufficiently small in mass and sfermions are light so that light sfermion 
exchange may enhance the LSP-LSP annihilation rates, or 
iv) there is a possibility of having s-channel Higgs exchanges 
where the exchanges may occur via CP-odd Higgs boson 
$A$ or via CP-even heavy (light) Higgs bosons $H$ ($h$) leading to the 
``funnel region'' of dark matter satisfied zone in 
the $m_{1/2}-m_0$ plane\cite{funnel}.    

Most recently XENON100\cite{xenon100May2010} experiment has published an updated result 
for the exclusion limit of $3 \times 10^{-8}~{\rm pb}$ for 
direct search of dark matter for a 
weakly interacting massive particle (WIMP) 
mass of 50~GeV at 90\% confidence 
limit. Additionally, the recent 5-tower result of the 
CDMS-II\cite{Ahmed:2009zw} experiment published a result of   
2 signal events with $0.6 \pm 0.1$ events as background. 
As mentioned by the authors, there is a significant 
chance that these two events might have been 
caused by background rather than by any real signal. 
However, a low recoil of 
12.3 to 15.5  
keV\cite{Ahmed:2009zw} 
possibly suggests a lighter WIMP with a mass not quite far from 
100 GeV.  
We may thus assume that 
the spin-independent 
direct detection cross section limit ($\sigma^{SI}_{{\tilde \chi}_1^0p}$) 
set by the experiment, namely $ 3.8 \times 10^{-8}$ $\rm pb$ at a WIMP 
mass in the vicinity of 70~GeV may become an interesting zone in the near 
future in a light LSP scenario. Several analyses 
have used this theme in the recent past\cite{CDMSlightWIMP} 
and there have been several recent works\cite{CDMSotherSUSY} in SUSY models in the context of the 
CDMS-II result. 
For practical purposes, we will probe the region of $5~\times 10^{-9}$ to 
$5~\times 10^{-8}$ $\rm pb$ in a lighter LSP scenario (not too far 
from 100 GeV) that 
satisfies the following relic density bounds from WMAP\cite{WMAPdata}:  
\begin{equation}
\Omega_{{\tilde \chi}_1^0} h^2 = 0.1099 \ \ \pm 0.0186~(3\sigma).
\label{relicdensity}
\end{equation}
A possibly preferred scenario where the LSP is light and at the same time 
is dominated by Bino may be achieved if neutralino relic density 
is brought to an acceptable level (from a typical over-abundance) 
via a light Higgs pole annihilation mechanism.    
Computation of relic density 
in mSUGRA with nonvanishing trilinear coupling for large $\tan\beta$ 
showed such a possibility in Ref.\cite{djouadiLightHiggs} that also imposed 
constraints from flavor physics like that from 
$b \rightarrow s  \gamma$ and $B_s \rightarrow \mu^+ \mu^-$.   
Here in this work, we will 
particularly explore the reach of the direct detection cross section  
in relation to any small recoil data of CDMS-II 
in such a scenario with light Higgs pole annihilation. 
In addition to imposing the LEP2 limit\cite{hlim} for $m_h$, 
the mass of the light Higgs boson, we will also explore 
the parameter space in regard to the constraints like  
$b \rightarrow s  \gamma$ and $B_s \rightarrow \mu^+ \mu^-$ 
processes. We will particularly 
discuss the effect of considering nonminimal flavor 
violating (non-MFV) scenarios with a wider point of view 
that may be considered without affecting 
much the spectra or the dark matter related results.  
We will further probe the possible signatures coming out of 
LHC in regard to MSSM Higgs bosons and weakly interacting 
gauginos $( \tilde\chi^0_2, \tilde\chi^\pm_1) $ in the dark matter 
satisfied zone with larger direct detection cross section 
$\sigma^{SI}_{{\tilde \chi}_1^0p}$. We note that LHC may help to probe 
the mass of the lighter Higgs boson ($m_h$) from 
$pp \to h \to \gamma \gamma$ mode, whereas an 
existence of light Higgs pole annihilation would indicate a  
light LSP scenario with a value of $\mlspone$ that would 
satisfy $2\mlspone \lsim m_h$. 
Additionally, at the LHC,
one would expect to see a signal of three leptons and a missing 
transverse energy namely $3 \ell + \ptm$ mainly
from the pair production of $\tilde\chi^0_2 \tilde\chi^\pm_1$
followed by their decays via leptonic modes,
$pp \to \tilde\chi^0_2\tilde\chi^\pm_1 \to 3\ell + \ptm $ as a classic
signature of SUSY.
We also note that in regard to the decay 
$\tilde\chi^0_2 \to \ell^+ \ell^-            
\tilde\chi^0_1 $, the endpoint of {\it opposite sign same flavor} (OSSF)
dilepton invariant mass distribution  $m_{\ell^{+}\ell^{-}}$ leads to a
good determination of the mass difference 
$m_{\tilde\chi^0_2} - m_{\tilde\chi^0_1}$\cite{ATLAS,invariantmass}. 
  
This article is organized as follows. In Sec.~2, we discuss
the issues of direct detection of dark matter particularly 
the spin-independent cross section $\sigma^{SI}_{{\tilde \chi}_1^0p}$ 
in relation to the parameter space of mSUGRA that satisfies the WMAP data 
via s-channel Higgs pole annihilation. In Sec.~3, we identify 
the preferred sign of the trilinear coupling in relation to 
the above s-channel annihilation.  In Sec.~4, we discuss
the implications of WMAP data and low energy constraints on the
mSUGRA parameter space that is interesting for 
a CDMS-II type of event with a small LSP mass. 
In Sec.~5, we discuss several possible signatures of this scenario at the LHC.
We also comment on how the correlations between
different LHC signatures and the direct detection of dark matter
may help determine the masses of $\tilde\chi^0_1$ and $\tilde\chi^0_2$.
Finally, we summarise our findings in Sec.~6.
\section{Direct detection of dark matter} 
In general, the results of 
cross section of LSP annihilation to quarks can be used to
find the elastic scattering cross section of the LSP with quark,
thanks to crossing symmetry. Direct detections of
LSPs involve measurement of recoil energy of a nucleus due to
LSP-nucleon scattering\cite{DMreview,ellis2000,alldirectdetect}. 
Typical detector materials are scintillators like
NaI, semiconductors like Ge and noble liquids like Xe.
The elastic scattering of neutralino with nucleons are divided into two
types (i) spin-independent (SI): a neutralino coherently interacts with the
nucleus and (ii) spin-dependent (SD):  a neutralino interacts with
matter via axial vector coupling. The effective Lagrangian that describes $\chi-q$ 
elastic scattering for
a small velocity is given by,
\begin{equation}
{\cal L} = \alpha'_{qi}\bar{\chi} \gamma^\mu \gamma^5 \chi \bar{q_{i}}
\gamma_{\mu} \gamma^{5} q_{i} +
\alpha_{qi} \bar{\chi} \chi \bar{q_{i}} q_{i}~.
\label{lagxsection}
\end{equation}
Here, the first term represents spin-dependent scattering
while the second term refers to spin-independent scattering.
The terms assume summing over the quark
flavors $q$ as well as that for up and down type of quarks
(for $i=1$ and $i=2$ respectively).
The neutralino-quark coupling coefficients
$\alpha_{q}$ and $\alpha'_{q}$ contain all SUSY model dependent 
information\cite{ellis2000}.
The spin-independent scattering cross section of a neutralino with a target
nucleus of proton number (atomic number) 
$Z$ and neutron number $A-Z$ ($A$ being the
mass number) is given by,
\begin{equation}
\sigma^{SI} = \frac{4 m_{r}^{2}}{\pi} \left[ Z f_{p} + (A-Z) f_{n}
\right]^{2}~.
\label{sitotal}
\end{equation}
Here, $m_r$ is the reduced mass
defined by $m_r=\frac{m_\chi m_N}{(m_\chi+ m_N)}$, where
$m_N$ refers to the mass of the nucleus.
The quantities $f_p$ and $f_n$
contain all the information of short-distance physics and nuclear
partonic strengths. These are given by,
\begin{equation}
\frac{f_{p, (n)}}{m_{p, (n)}} = \sum_{q=u, d, s} f_{Tq}^{(p, (n))} 
\frac{\alpha_{q}}{m_{q}} +
\frac{2}{27} f_{TG}^{(p, (n))} \sum_{c, b, t} \frac{\alpha_{q}}{m_q}~. 
\label{fpn}
\end{equation}
$f_{Tq}^{(p, (n))}$ and $ f_{TG}^{(p, (n))}$ refer to 
interactions of neutralino with quarks and gluons (via quark/squark
loop diagrams) respectively. For $f_{Tq}^{(p, (n))}$ one has, 
\begin{equation}
m_{p, (n)} f_{Tq}^{(p, (n))} = \langle p, (n) | m_{q} \bar{q} q | p, (n) 
\rangle \equiv 
m_q B_q~. 
\label{defbq}
\end{equation}
The quantities $B_q$ may be computed from a few hadronic mass data. 
The gluon related part namely $f_{TG}^{(p, (n))}$ is given 
by\cite{ellis2000,alldirectdetect}, 
\begin{equation}
 f_{TG}^{(p, (n))} = 1 - \sum_{q=u, d, s} f_{Tq}^{(p, (n))}~. 
\end{equation}
The numerical values of $f_{Tq}^{(p, (n))}$ 
may be seen in Ref.\cite{ellis2000}. 
We compute $\sigma^{SI}$ by using the code DarkSusy\cite{darksusy}.
In contrast to $\sigma^{SI}$, the spin-dependent cross section denoted by 
$\sigma^{SD}$   
(for scattering of LSP with the target nucleus) 
does not depend on $A$ or $Z$, rather it scales with $J(J+1)$ where 
$J$ is the total nuclear spin.
In general, the spin-independent
neutralino-nucleon scattering cross sections
(where $\sigma_{\chi p}^{SI} \simeq \sigma_{\chi n}^{SI}$) are
appreciably smaller than the corresponding spin-dependent
cross sections ($\sigma_{\chi p}^{SD} \simeq              
\sigma_{\chi n}^{SD}$).  However considering the fact that 
$\sigma^{SD} \propto J(J+1)$ and  
$\sigma^{SI} \propto Z^2, (A-Z)^2$ one finds that 
$\sigma^{SI}$ to be considerably larger for moderately heavy
elements ($A>30$)
like Xe, Ge etc. The experiments like EDELWEISS\cite{edelweiss} and 
CDMS\cite{CDMSlightWIMP} 
use natural germanium (almost purely spinless) as the target 
material and these hardly have any 
spin-sensitivities\cite{Bednyakov:2008gv}. 
   It is important to know the composition of 
the $\lspone$ in terms of Bino, Wino and Higgsinos in MSSM while analyzing  
the WMAP satisfied relic density region as well as the 
LSP-nucleon scattering cross section for a direct detection.
The amount of Higgsino mixing in a typically Bino-dominated LSP 
scenario of mSUGRA depends on the relative value of $|\mu|$ and 
$m_1$ (the mass of Bino). 
The scalar cross section depends on t-channel Higgs exchange
($h,H$) and s-channel squark exchange diagrams. Unless, the squark masses 
are close to the mass of the LSP, the Higgs exchange diagrams 
usually dominate over the s-channel diagrams\cite{Drees:1993bu}. 
We focus on a decoupling region of Higgs\cite{decouplingHiggs}, so that there is a considerable difference of masses between $h$ and $H$-bosons.
We also focus on the case where LSP is principally composed of Bino,  
$\tan\beta$ is large and there is a resonance annihilation via 
$h$-boson that is consistent with the relic density from the 
WMAP limits. Following the results of Refs.\cite{Drees:1993bu,hisanoNew}
we see that both the contributions involving  
lighter as well as heavier Higgs bosons in the direct detection are 
important. 
Indeed in the characteristic zone of the parameter space 
as mentioned above, the heavier Higgs boson contribution becomes 
dominant in comparison to the contribution from the 
lighter Higgs boson\cite{Belanger:2004ag}. This holds in spite of the 
fact that $m_H$ can be larger by a factor of 3 to 4 compared to $m_h$.
In this light Higgs annihilation zone we will see that $\sigma^{SI}_{{\tilde \chi}_1^0p}$ can be large enough to reach the CDMS-II limit.    
 
\section{Preference of negative $A_0$ for light Higgs-pole resonance} In this analysis we 
especially investigate the parameter space that satisfies 
the relic density limits from WMAP data via LSP annihilations through light Higgs resonance. This necessitates a closer study of 
the region with small $m_{1/2}$ near its domain just allowed by the LEP2 limit 
of lighter chargino mass $m_{{\tilde \chi}_1^\pm}$\cite{limits}. 
We will see how the LEP2 constraint affects the lowest 
possible value of $m_{1/2}$ in a nonvanishing $A_0$ scenario\footnote{
We have imposed restrictions to avoid the appearance of 
charge and color breaking (CCB) minima\cite{casasCCB} 
in the parameter space.}. 
This in turn requires a discussion on the behaviors of 
i) $\mu$ and ii) $m_2$\footnote{We consider two-loop RGEs along with 
threshold corrections.} for a variation of $A_0$.
We note that in mSUGRA the variation of $\mu$ with $A_0$ 
has the following pattern. For $A_0<0$, $|\mu|$ 
increases with $|A_0|$, whereas $|\mu|$ increases with $A_0$ as long 
as $A_0 \geq m_{1/2}$. On the other hand,  
$\mu$ is near its minimum when one has $A_0 \lsim m_{1/2}$.
As a result, one finds that for a given set of $m_0$, $m_{1/2}$ and $\tan\beta$ a change of $A_0$
from $A_0=x$ to $A_0=-x$ where $x\sim \mathcal{O} ~{\rm TeV}$ causes  
$|\mu|$ to vary from a large to a larger value. 
This in turn causes ${\tilde \chi}_1^\pm$ to be dominated   
by Wino in the region with a smaller $m_{1/2}$, the amount of 
Wino content is larger for the case of $A_0=-x$ in comparison to 
that of $A_0=x$ where the same content is still 
significantly large. Thus, while probing 
the region of small gaugino masses with Wino-dominated 
${\tilde \chi}_1^\pm$, it is important to examine the variation of
$m_2$ with nonvanishing $A_0$.  
In this context, Table~\ref{tabgauginomasses} shows the masses of 
electroweak gauginos, the value of higgsino mass parameter, 
the masses of lighter chargino, lighter neutralino and h-boson  
for $\tan\beta=50$, $m_{1/2}=150$~GeV, $m_0=750$~GeV and $\mu>0$ for 
$A_0=\pm 1250$~GeV. This will indeed demonstrate the effect of two-loop RGEs 
along with threshold corrections on the electroweak gaugino masses. 
Clearly, both $m_1$ and $m_2$ are smaller for $A_0=x$ in 
comparison to the same quantities corresponding to $A_0=-x$.
On the other hand, the values of $\mu$ for both signs of $A_0$ 
are sufficiently large so that  the lighter chargino as well as the lighter neutralino 
are both highly gaugino-dominated. The value of $m_h$ is smaller for $A_0>0$. 
Thus, as we decrease $m_{1/2}$, the LEP2 bound of chargino mass is reached 
at a relatively larger $m_{1/2}$ for $A_0=x$ in comparison to the case of 
$A_0=-x$. Similarly,  the LEP2 limit of $m_h$ is reached at a relatively 
larger $m_{1/2}$ for $A_0=x$ than the case of $A_0=-x$. 
The latter favors 
larger loop corrections from the top-stop sector\cite{Carena:2002es}. 
  
In brief, imposition of LEP2 bounds of $m_{{\tilde \chi}_1^\pm}$ and $m_h$ 
(considering also the aforesaid 3~GeV theoretical uncertainty) 
would allow us to reach to a smaller $m_{1/2}$ value for a negative $A_0$ than for a 
positive $A_0$\cite{Chattopadhyay:2007di}. 
As a result, Bino can be light enough to satisfy a 
light Higgs-pole annihilation condition $2\mlspone \lsim m_h$ while obeying the 
WMAP limits in a LEP2 satisfied parameter zone. Such a light LSP would 
undoubtedly be 
interesting in view of the CDMS-II results with two possible events.
\begin{table}[!ht]
\begin{center}
\begin{tabular}[ht]{|c|c|c|c|c|c|c|c|c|c|c|c|}
\hline
\multicolumn{6}{|c|}{$A_0=1250.0$~GeV} & \multicolumn{6}{|c|}{$A_0=-1250.0$~GeV}\\
\hline
$m_1$ & $m_2$ & $\mu$ & $m_{{\tilde \chi}_1^\pm}$ & $m_{{\tilde \chi}_1^0}$ & $m_h$ & 
$m_1$ & $m_2$ & $\mu$ & $m_{{\tilde \chi}_1^\pm}$ & $m_{{\tilde \chi}_1^0}$ & $m_h$ \\
\hline
58.8  & 111.2 & 352.4 & 104.7 &  57.5 & 106.6 &
62.4  & 118.7 & 492.3 & 121.9 &  62.2 & 114.3\\

\hline
\end{tabular}
\caption{Comparison of spectra electroweak gaugino masses, higgsino masses, 
the masses of lighter chargino, lighter neutralino and h-boson  
for $\tan\beta=50$, $m_{1/2}=150$~GeV, 
$m_0=750$~GeV and $\mu>0$ for $A_0=\pm 1250$~GeV. This is given as an  
example to demonstrate the effect of two-loop RGEs along with 
threshold corrections on the electroweak gaugino masses which are smaller 
for $A_0>0$. These are of course not valid parameter points 
satisfying WMAP data for neutralino relic density. }
\label{tabgauginomasses}
\end{center}
\end{table}

\begin{figure}[htbp]
\begin{center}
\includegraphics[width=10cm]{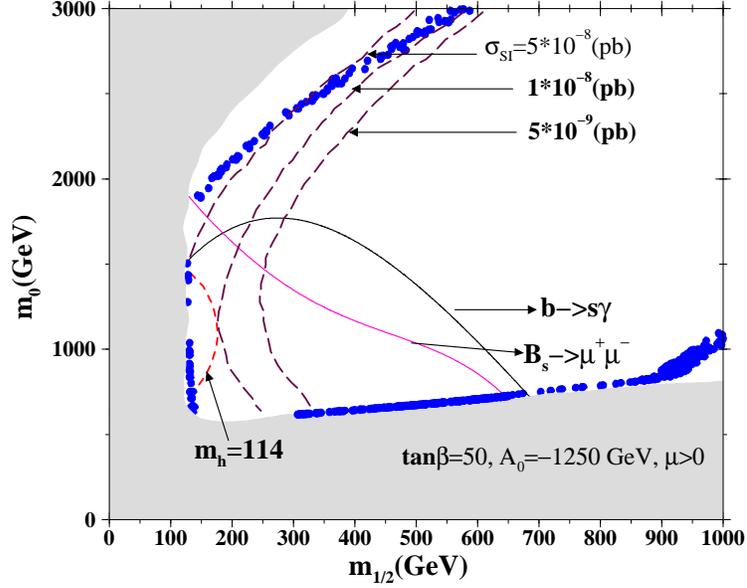}
\caption{Results for $\tan\beta=50$ and $A_0=-1250 $~GeV for $\mu>0$. 
The WMAP satisfied zones in $m_0-m_{1/2}$ plane are shown in blue. 
The lower blue region with $m_{1/2}<900$~GeV and $m_0<1$~TeV satisfies 
WMAP data via LSP-stau coannihilation. The extension of this blue 
region in the larger $m_{1/2}$ side is associated with LSP-LSP resonance annihilation via $A/H$-bosons. 
The left-most vertical blue region with small $m_{1/2}<150$~GeV 
corresponds to $s$-channel annihilation via light Higgs boson $h$.
Contours shown for $\sigma_{SI}^{{\tilde \chi}p}$ from $5\times 10^{-9}$~pb 
to $5\times 10^{-8}$~pb. 
The left hand sides of the    
contours for $Br(b \rightarrow s\gamma)$ and 
$ B_s \rightarrow \mu^+\mu^-$ are discarded in the 
minimal flavor violation scenario. 
We have considered the possibility of a general flavor violation scenario as 
mentioned in the text. The details of the discarded regions are mentioned in the text.
}
\label{figA0negative}
\end{center}
\end{figure}
\section{WMAP data and low energy constraints} 
We now explain the results of imposing dark matter constraint for 
$A_0<0$ for large $\tan\beta$ and the relevance of the light-Higgs pole 
mediated annihilations in relation to the CDMS-II results.
Fig.\ref{figA0negative} shows the results for $\mu>0$ in $m_0-m_{1/2}$ 
plane for 
$\tan\beta=50$ and $A_0=-1250$~GeV{\footnote {We consider 
top pole mass $m_t = 171.4$~GeV. }}. 
Almost the entire parameter space has a highly 
Bino-dominated LSP except for the region where $m_0$ is large. 
The latter region is associated with a significant amount of higgsinos in the 
composition of the LSP. 
The WMAP constraint of 
Eq.\ref{relicdensity} is satisfied in the blue shaded regions which are 
divided into three broad zones. The lower blue region with 
$m_{1/2}<900$~GeV and $m_0<1$~TeV satisfies 
WMAP data via LSP-stau coannihilation. The extension of this blue 
region in the 
larger $m_{1/2}$ side is associated with LSP-LSP resonance annihilation 
via $A/H$-bosons. 
The upper blue $m_0$ region satisfying the WMAP limits 
is characterised by a considerably large Higgsino fraction in the 
composition of LSP. Here the $s$-channel annihilation occurs via 
$A/H$-bosons that may be several widths ($\Gamma_{A,H}$) away.  
The left-most blue region with small $m_{1/2}$ corresponds to 
$s$-channel annihilation via light Higgs boson $h$. Bracketing 
the sensitivity level of the CDMS-II result in particular 
$3.8\times 10^{-8}$~pb or the XENON100 data of $3\times 10^{-8}$~pb we 
have drawn contours of the spin-independent ${\tilde \chi}p$ cross section
$\sigma_{SI}^{{\tilde \chi}p}$ from $5\times 10^{-9}$~pb 
to $5\times 10^{-8}$~pb. We see that apart from the large $m_0$ region 
satisfying the WMAP data via $A/H$ resonance,  
there exists a significant possibility to probe the region of 
light Higgs boson resonance that produces a large 
$\sigma_{SI}^{{\tilde \chi}p}$. 
We also show a contour of $m_h=114$~GeV which is close to the LEP2 bound 
of $h$ of 114.4~GeV. We note that although we have considered 
a large value for $\tan\beta$ the pseudoscalar Higgs boson 
is not very light.  Thus the latter being still in the 
decoupling region the associated 
lighter Higgs boson $h$ is standard model like. However, we should also 
note that there is about a 3~GeV uncertainty in the theoretical result 
of $m_h$ while including the loop corrections\cite{higgsuncertainty}. 
Hence we consider 
a value of $m_h=111$~GeV to be an effective lower limit. 
We further draw contours corresponding to the 
limits from $Br(b \rightarrow s\gamma)$\cite{bsg-sm,bsg-bsm,bsg-recent}
 and 
$Br(B_s \rightarrow \mu^+\mu^-)$\cite{bsmumurefs,bsmumuCDF}. The limits for $Br(b \rightarrow s\gamma)$ 
at 3$\sigma$ level is given by\cite{bsg-recent} 
\begin{equation}
2.77\times10^{-4} <Br(b \rightarrow s\gamma) <4.33\times10^{-4}.
\end{equation}
Clearly, the above limits of $ b \rightarrow s\gamma$ 
are not favorable to the light 
Higgs pole annihilation region satisfying the WMAP constraint. 
We however point out that the above limits apply to models 
where perfect alignment of the squark and quark mass matrices 
are assumed. This indicates no extra mixing other than the 
Cabibbo-Kobayashi-Maskawa factors existing at the corresponding SM 
vertices. However, one may obtain a considerable degree of 
squark mixing at the electroweak scale if there is even a very small off-diagonal term 
at the grand unification scale. 
As discussed in Ref.\cite{Djouadi:2006be} this 
may considerably reduce the effect of 
the constraint from $b\rightarrow s \gamma$ on the parameter space of the SUSY model.
On the other hand, such a possible deviation from the minimal flavor violation (MFV)\cite{Colangelo:2008qp}
scenario does not cause any significant change in the sparticle masses or in the results of flavor 
conserving processes like neutralino annihilation.
Additionally, limits 
from $Br(B_s \rightarrow \mu^+\mu^-)$ becomes important in a 
large $\tan\beta$ and/or small pseudoscalar Higgs mass 
scenario since $Br(B_s \rightarrow \mu^+\mu^-)$
varies as $\tan^6\beta$ or in a light pseudoscalar Higgs boson scenario 
the same varies as $m_A^{-4}$\cite{bsmumurefs}. 
The present upper limit is given below\cite{bsmumuCDF}. 
\begin{equation}
Br(B_s \rightarrow \mu^+\mu^-) <5.8\times10^{-8}.
\end{equation}
We may as well consider a non-MFV scenario of squark mixing 
for $B_s \rightarrow \mu^+\mu^-$ and thereby consider 
the light Higgs pole annihilation region to be an open region for study.
References.\cite{nonMFV1,nonMFV2,Blazek:2010zz} 
may be seen for analyses with non-MFV scenarios 
for $Br(b \rightarrow s\gamma)$ and $ B_s \rightarrow \mu^+\mu^-$.  
Following the arguments and analyses of 
Refs.\cite{Djouadi:2006be,nonMFV1,nonMFV2,Blazek:2010zz}
we therefore prefer not to exclude any region allowed by the 
WMAP data on the basis of the above constraints from flavor physics.
The discarded regions are shown in gray. The bottom gray region is 
discarded via staus (${\tilde \tau}_1$) becoming the LSP. The left gray region 
with smaller $m_0$ with almost a vertical boundary is discarded via 
the LEP2 chargino mass bound requirement $m_{\tilde \chi^\pm}>103.5 \rm ~GeV$. 
The top left gray region is discarded because of absence of radiative electroweak symmetry breaking.

Finally for completeness we show 
Fig.\ref{figA0zero} and Fig.\ref{figA0positive} in regard to the 
results corresponding to 
a vanishing and a positive ($=1250$~GeV) $A_0$ respectively.  
Unlike Fig.\ref{figA0negative} these two cases do not have any light Higgs 
pole 
annihilation region. The s-channel annihilations occur only via $A/H$ 
bosons for these two cases. 
   
\begin{figure}[!h]
\vspace*{-0.05in}
\mygraph{figA0zero}{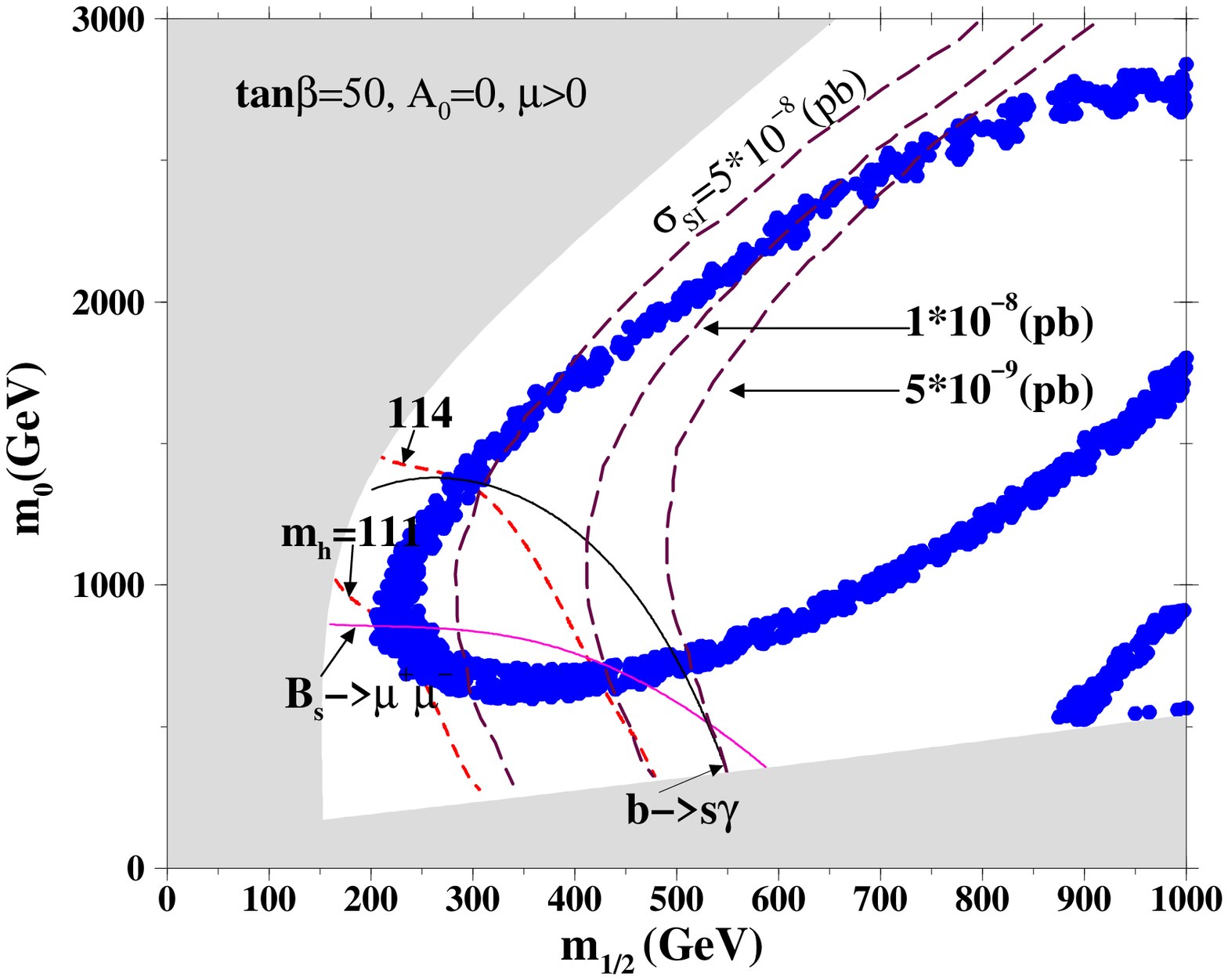}
\hspace*{0.5in}
\mygraph{figA0positive}{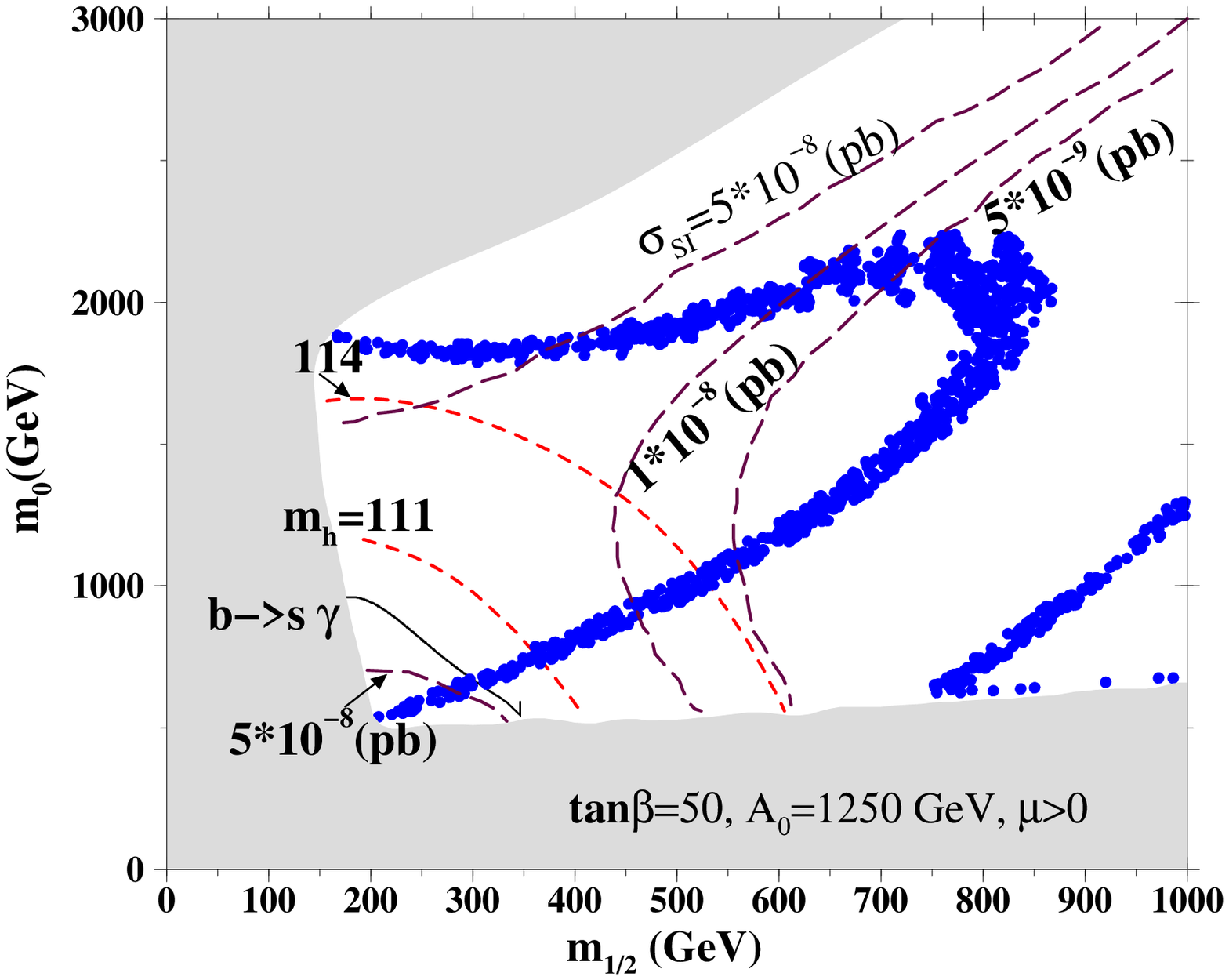}
\caption{
Similar to Fig.\ref{figA0negative} except for $A_0=0$ and $A_0=1250$~GeV as shown above. There are no light Higgs resonance regions in these two cases.}
\label{zeroAndpositive}
\end{figure}

\section{LHC phenomenology}
We now qualitatively discuss a few issues regarding 
various LHC searches for the Higgs bosons as well as strongly and 
weakly interacting gauginos in our scenario where LSP-LSP annihilation 
mechanism related to dark matter relics 
occurs via the light Higgs resonance channel. We present two benchmark points 
namely A and B in Table~\ref{tabfunnel} with the corresponding 
SUSY spectra in this regard.  
Detailed analyses for the searches 
will be reported in Ref.\cite{uc_dd_dg_mm}.  The benchmark points satisfy  
the necessary phenomenological constraints including the LEP2 bound 
for the Higgs boson mass except the fact that they do not 
satisfy the limits from $Br(b \rightarrow s\gamma)$ and $ B_s \rightarrow \mu^+\mu^-$, if minimal flavor violation is assumed. As mentioned before, the 
above flavor constraints are relaxed in a general or a non-MFV type of
estimation. 
The leading decay modes along with the branching ratios (BR) 
corresponding to the points A and B are given in Table~\ref{tab:msugra_br}. 
We note that the above points obey the  
decoupling limit $(m_A >> M_Z)$ so that the lighter Higgs boson $h$ behaves 
like a standard model like Higgs boson. 
We see that WMAP satisfied zone corresponding to 
the light Higgs pole annihilation 
is very close to the exclusion limit 
provided by the LEP2 and Tevatron. Being in a 
decoupling zone, the pseudoscalar Higgs boson 
($A$) is appreciably massive with also having a large 
width ($\Gamma_A$).  
Thus it can hardly have a simultaneous contribution to 
the resonance annihilation in a parameter zone where there is 
a light Higgs boson in the s-channel. 
On the other hand 
with a small width $\Gamma_h$ for the h-boson, 
the LSP mass $m_{\chi_1^0}$ has a tight 
constraint, namely $2 m_{\chi_1^0} \simeq m_h$ from the 
WMAP data. 
This particular
feature when coupled with different observables related to the accelerator
physics may provide a very useful estimate for the supersymmetric 
mass spectrum directly from the experimental data.

\begin{table}[!ht]
\begin{center}
\begin{tabular}[ht]{|l|c|c|}
\hline
parameters & A & B \\
\hline
$\tan\beta$ &50.0  &50.0\\
$m_{1/2}$ &132.0  &130.0 \\
$m_0$ &750.0  &915.0 \\
$A_0$ & -1250.0  & -1250. \\
$sgn(\mu)$ &1  &1 \\
\hline
\hline
$\mu$ & 476.63 & 461.94  \\
$m_{\tg}$ & 384.15 &   389.54   \\
$m_{\tu_L}$ & 793.86 &  944.17  \\
$m_{\tst_1}$ &  286.49 & 425.17   \\
$m_{\tst_2}$ &  550.38  & 638.94  \\
$m_{\tb_1}$ & 453.97 & 570.91   \\
$m_{\tb_2}$ &  592.56 & 701.47  \\
$m_{\te_L}$ & 753.03 & 915.73   \\
$m_{{\tilde \tau}_1}$ & 373.21 & 539.54   \\
$m_{{\wt\chi_1}^{\pm}}$ & 107.51 &  106.72  \\
$m_{{\wt\chi_2}^{\pm}}$ & 486.15 & 474.13 \\
$\mlspfour$ & 483.47 & 471.43  \\
$\mlspthree$ & 479.13 & 466.90   \\ 
$\mlsptwo$ & 107.54 &106.75 \\ 
$\mlspone$ & 54.67 & 54.15   \\ 
$m_A$ & 336.09 &  352.60   \\
$m_{H^+}$ &  346.84 &  363.02   \\
$m_h$ & 113.87 & 113.0 \\
$\Omega_{\lspone}h^2$& 0.120 & 0.114  \\
$\sigma_{SI}(pb)$& 2.75$\times$ $10^{-8}$ & 2.26$\times 10^{-8}$\\
\hline
\end{tabular}
\caption{Benchmark points A and B along with the spectra. Masses are shown 
in~GeV.}
\label{tabfunnel}

\end{center}
\end{table}


\begin{table}[!ht]
\begin{center}\
\begin{tabular}{|c|c|c|}
       \hline
Decay modes & A &B\\
(squark/gluino) & &\\
\hline
\hline
 $\widetilde g \rightarrow \lsptwo b {\bar b} $&44.0  &36.0\\
\hline
 $\widetilde g \rightarrow \chonem t b$&14.0  &12.0\\
\hline
$\widetilde g \rightarrow \chonem u d$ &18.0  &24.0\\
\hline
\hline
$\widetilde b_1 \rightarrow \widetilde g b$&38.0  &66.0\\
\hline
$\widetilde b_1 \rightarrow \lsptwo b$ &17.0  &10.0\\
\hline
$\widetilde b_1 \rightarrow \chonem t$ &23.0  &15.40\\
\hline
$\widetilde b_1 \rightarrow \lstop W^{-}$ &17.0  &5.30\\
\hline
\hline
$\lstop \rightarrow \chonep b$ &72.40  &53.70\\
\hline
$\lstop \rightarrow \lspone t$ &22.0  &29.40\\
\hline
$\lstop \rightarrow \lsptwo t$ &5.0  &16.80\\
\hline
\hline
$\lsptwo \rightarrow \lspone q \bar q$ &48.0&50.0\\
\hline
$\lsptwo \rightarrow \lspone b \bar b$ &33.0&28.60\\
\hline
$\lsptwo \rightarrow \lspone \ell^+ \ell^-$, $(\ell = e, \mu)$ &2.54&3.7\\
\hline
$\lsptwo \rightarrow \lspone \tau^+ \tau^-$ &8.7&4.56\\
\hline
$\lsptwo \rightarrow \lspone \nu \bar \nu$ &6.0&10.0\\
\hline
$\chonep \rightarrow \lspone u {\bar d} $ &69.0&68.60\\
\hline
$\chonep \rightarrow \lspone \ell^+ \nu_{\ell} $ &20&20.0\\
\hline
$\chonep \rightarrow \lspone \tau^+ \nu_{\tau} $ &11.0&10.0\\
\hline
\end{tabular}

\end{center}
\caption{ The branching ratios($\%$)of the dominant decay modes of the 
gluinos, squarks and lighter electroweak gauginos in mSUGRA for the points
 A and B.}
\label{tab:msugra_br}
\end{table}
Before going into details of the 
LHC signatures predicted by this scenario, let us first discuss few 
salient features of the low energy SUSY spectrum which can be inferred 
from the Table~\ref{tabfunnel}.
\begin{itemize} 
\item Large $\tan\beta$.
\item The Higgs sector falls within the decoupling regime, with 
moderately heavy $H$ and $A$, and light SM like Higgs boson $h$. 
\item Light LSP mass, $\mlspone \approx m_h/2 $, while 
lighter chargino $(\tilde \chi^\pm_1)$ and second lightest neutralino
$(\tilde \chi^0_2)$ are almost degenerate with masses $\sim 105 $ GeV.
\item Heavy first two generations of squarks and sleptons with masses above 
700 GeV.
\item Light gluinos $(m_{\tilde g} \leq 400~{\rm GeV} ) $
and lighter third generation of squarks and sleptons.
\end{itemize} 
We start our discussions with the light MSSM Higgs boson $h$, which 
as mentioned before behaves like a SM Higgs boson.
The dominant production mechanism of 
$h$ is via $gg \to h $ process followed by (i) $qq h$ production through
vector boson fusion, (ii) associated productions with $W^\pm$ or $Z$ bosons,
and (iii) in association with a $t {\bar t}$ pair.
The $h \to \gamma \gamma $ decay mode is the best channel to look
for a light Higgs boson $h$ channel with $m_h < 130 $ GeV \cite{ATLAS,CMS}. 
For the Higgs mass of our interest ($\sim 115$ GeV), one has 
the $\sigma (pp\to h)_{\rm inclusive} \times {\rm BR} \sim {\cal O}(100)$ fb 
and for a $5\sigma$ discovery one requires integrated luminosity of 
$30~{\rm fb}^{-1}$ at the 14 TeV LHC run \cite{CMS}. 

We now turn our focus on the production of heavy neutral MSSM Higgs bosons.
The bottom quark Yukawa couplings to both $H$ and $A$
become strongly enhanced at large $\tan\beta$. At the LHC,
the production process of $\Phi$, ($\Phi \equiv H,A $) 
in association with bottom quark benefits from a huge enhancement 
factor of $\tan^2\beta$ compared to the SM case.
Motivated by this large enhancement of the cross section, several analyses 
have been performed to study the production process of $\Phi$
either from bottom quark fusion $b {\bar b} \to \Phi $ 
\cite{Dicus1,Dicus2,Balazs,Maltoni,Harlander}, or in association with
one or two $b$-quarks with high transverse momentum $(p_T)$. 
These authors 
\cite{Choudhury,Huang,campbell,Cao,Dawson:2007ur} studied 
$b g \to b \Phi $ process both in the context of Tevatron and the LHC. 
On the other hand, $\Phi$ can be produced in association with two high 
$p_T$ $b$-quarks through the leading order sub process 
$gg \to b {\bar b} \Phi $ 
\cite{Dicus1,hbbmm,Plumper,Dittmaier,Dawson:2003kb}.
However, the above process turned out to be less
promising than Higgs production with a single high $p_T$ b-quark at the
LHC \cite{campbell}. It has been shown that the discovery of heavy Higgs
bosons through $bg \to b \Phi \to b \tau^+\tau^-$ mode at large $\tan\beta $ 
is very promising at the LHC \cite{kunszt,richter-was,kao-dicus}.
The Higgs mass can be 
reconstructed in the $\Phi \to \tau^+ \tau^- $ channel from the momenta
of the visible decay products of tau,  {\em i.e.}, 
either leptons or jets and the missing  
transverse energy $(\ptm)$ 
(arising from escaping neutrinos in the $\tau$ decay),
assuming that neutrinos are highly collinear due to the large boost of the
$\tau$'s \cite{CMS}. Very recently another possibility of discovering
$\Phi$ at the LHC with $b$-quarks has been proposed: 
$bg \to b\Phi\to b (b {\bar b})$ leading to $3b$ final state
\cite{kao-shankar}.
Following the above discussions it is worth noting that the
LHC will be able to probe the full range of $H/A$ masses given by 
the part of the mSUGRA parameter space that is consistent with 
the WMAP data satisfied via light Higgs resonance 
annihilation and associated with 
a large direct detection cross section 
compatible with the possible CDMS-II events.

At the LHC, multilepton + missing transverse energy final states with 
very little hadronic activity is one of the most promising discovery channels 
of supersymmetry. Such final states may come from the leptonic decay of the
pair of heavy gauginos (such as $\tilde\chi^\pm_1,\tilde\chi^0_2 $) 
through real or virtual $W^\pm, Z^0$ or via decays of sleptons to leptons 
and a pair of 
LSPs (main contributor to missing transverse energy). The electroweak 
production of heavy gaugino pairs  
\begin{eqnarray}
pp &\to &\tilde\chi^0_2 \tilde \chi^\pm_1~~, \nonumber \\
\tilde\chi^0_2 &\to &\tilde \chi^0_1  \ell^+ \ell^-~;~ 
\tilde\chi^\pm_1 \to \tilde \chi^0_1  \ell^\pm \nu_\ell \ ,~~~~
\ell = e~{\rm or}~\mu 
\label{n2c1_prod}
\end{eqnarray}
can lead to 
$3 \ell + \ptm $ signal. 
The $3\ell + \ptm $ channel may also get some contribution
from the production of heavier gauginos 
$ pp \to \tilde \chi^\pm_1 \tilde \chi^0_3, \tilde \chi^\pm_1 \tilde \chi^0_4, 
\tilde \chi^\pm_2 \tilde \chi^0_3, 
\tilde \chi^\pm_2 \tilde \chi^0_4 $
followed by their decays into leptonic final states. 
It is worth mentioning here, that, for our choice of benchmark points $(A,B)$,
the trilepton channels do not get any substantial contribution from these
heavier gauginos due to smaller production cross sections and also suppressed
branching ratios into leptonic final states. 
Hence, in our analysis, we only consider $3\ell^\pm + \ptm $ final state 
arising from Eq.(\ref{n2c1_prod}). 
In Table.\ref{tab:msugra_br}, we show the
branching ratios of $\tilde\chi^0_2 \to \tilde \chi^0_1 \ell^+ \ell^-$ 
and $\tilde\chi^+_1 \to \tilde \chi^0_1 \ell^+ \nu_\ell $ decays 
$(\ell = e,\mu)$. In Table \ref{tab:3lmpt}, we present the $\sigma ( pp \to 
\tilde \chi^0_2 \tilde\chi^\pm_1 \to 3\ell + \ptm) $ without 
any cuts for thw two benchmark points $(A,B)$ assuming three possible values of
$\sqrt{s} = 7, 12 $ and 14 TeV at the LHC. 

\begin{table}
\begin{center}
\footnotesize
\begin{tabular}{|c|c|c|c|}
\hline
Benchmark &
\multicolumn{3}{c|}{$\sqrt{s}$ (TeV) } \\
\cline{2-4}
points & $ 7 $  & $ 12 $ & $ 14 $ \\
\hline
$A $ & 21.3 & 44.9 &  54.9 \\
\hline
$B $ &33.8 &71.3 & 87.1  
\\
\hline
\end{tabular}
\end{center}
\caption{Leading order cross sections (fb) 
for $pp \to \tilde\chi^0_2 \tilde \chi^\pm_1 \to
3\ell + \ptm $ for two benchmark points.}
\label{tab:3lmpt}
\end{table}
 In this trilepton signal i.e., $3\ell + \ptm$ channel, 
any choice of the three charged leptons ($l=$ $e$ or $\mu$) must consist 
of an OSSF pair and an additional lepton. 
The end point of the OSSF dilepton 
invariant mass $(m_{\ell^+\ell^-})$ distribution leads to a good 
determination of the mass difference $\Delta m \equiv 
m_{\tilde\chi^0_2}- m_{\tilde\chi^0_1}$ \cite{invariantmass,ATLAS}. 

On the other hand, after reconstruction of 
the light Higgs boson mass $m_h$
from $h$-decays and assuming $h$-pole annihilation as the 
mechanism for satisfying the WMAP data, it is possible to estimate the mass
$m_{\tilde\chi^0_1}$, which is $m_{\tilde\chi^0_1} \sim m_h/2$.
Then the mass of $\tilde\chi^0_2$ may be estimated by 
combining $m_{\tilde\chi^0_1}$ and with the edge in the 
$m_{\ell^+ \ell^-}$ distribution.
It may be recalled that for the present choice of
benchmark points, $\tilde\chi^0_1$ 
is completely Bino-dominated, while $\tilde\chi^0_2$ is Wino-dominated.
This leads to one to one correspondence between $U(1)$ soft gaugino mass 
parameter $m_1$ with $m_{\tilde\chi^0_1}$ and $SU(2)$ soft gaugino mass 
parameter $m_2$ with $m_{\tilde\chi^0_2}$. Hence, we expect that the knowledge 
of both $m_{\tilde\chi^0_1}$ and $m_{\tilde\chi^0_2}$ may be used to determine
the mass parameters $m_1 $ and $m_2$. It would be interesting 
to study whether the mass pattern of $m_1 $ and $m_2$ could hint towards
the gaugino mass unification which model like mSUGRA assumes.


Finally, we conclude this section, by highlighting following important 
observations:
\begin{enumerate}
\item The light Higgs boson mass can be determined with a reasonably
good precision via $h\to \gamma \gamma $ decay mode. 
\item In the present scenario, where $\tan\beta $ is large, 
both neutral Higgs boson masses can be determined 
via $H/A \to \tau \tau $ decay.
\item Both $m_{\tilde \chi^0_1}$ 
and $m_{\tilde \chi^0_2}$ or equivalently $m_1$ and $m_2$ can be 
estimated by combining the data from the direct detection of 
dark matter and the LHC. 
\item As mentioned before, our choice of parameter space prefers light gluinos and
also lighter third generation sleptons and squarks. This naturally 
leads to large production rate for gluinos as well as lighter stop and 
sbottom squarks which in turn give rise to large number of 
multi $b$-jet  
final states in association with large missing energy both at 7 and 14 TeV LHC run. 
This will be another striking signature of this scenario.
\end{enumerate}

\section{Conclusions}
Recent experiments like XENON and CDMS-II provide us 
with exclusion limits within 
3-4 $\times 10^{-8}~pb$ for WIMP mass in the vicinity of 
50 to 70~GeV. CDMS-II results show a 2-signal events with 
$0.6 \pm 0.1$ events as background. It may be possible that these events 
did not come from any real signal. In spite of this, in case we do not 
see the events as just background, a low recoil of  
12.3 to 15.5 keV could possibly indicate a light WIMP detection 
whose mass could not be too far away than 100~GeV. A large direct detection 
cross section where LSP is light (and that also satisfies the WMAP data) 
in a generic model like mSUGRA can be possible in a parameter space 
with a large $\tan\beta$ and nonvanishing trilinear coupling $A_0$.   
The relic density constraint is satisfied via h-pole resonant annihilation 
mechanism. Regarding direct detection ${\tilde \chi}_1^0-p$ cross section 
which reaches the level of the CDMS-II events, both the 
contributions involving  
the lighter as well as the heavier Higgs bosons are 
important. In this large $\tan\beta$ ($=50$) 
based scenario with Bino-dominated 
${\tilde \chi}_1^0$, where light Higgs pole 
annihilation brings down the relic density to an acceptable 
level, the heavier Higgs boson contribution indeed dominates.  
This is in spite of the fact that $m_H$ can be larger by a 
factor of 3 to 4 compared to $m_h$.

Unlike the past analyses referred in this paper regarding the light 
Higgs-pole dominated LSP pair-annihilation processes for 
the cases with large values of $\tan\beta$, here we have worked on the direct 
detection prospects of the same region in relation to the recent interest 
of CDMS-II results.  We have further shown in adequate detail the preference 
of only one sign of $A_0$ that gives feasibility to obtain the desired zone 
of light-Higgs pole annihilation for dark matter.  
We particularly perform an open 
analysis by considering non-MFV assumptions 
in relation to the $b \rightarrow s \gamma $ and 
$B_s \rightarrow \mu^+ \mu^-$ constraints, unlike most other recent  
works.  We have considered the possibility of general flavor violation on 
$B_s \rightarrow \mu^+ \mu^-$, specifically keeping in mind about the 
recent analyses as referred in this work.
We believe that unless the 
MFV-based limits from flavor physics are violated by a very large extent,  
it is important not to discard appropriate scenarios that keep 
the region with the light Higgs-pole dominated 
pair-annihilation active. This is particularly relevant   
because of the ongoing and upcoming precision direct detection 
dark matter experiments and the advantage of having an associated 
light Higgs spectra as well as a relatively light overall spectra. 
In the last half of the paper we present a
qualitative discussions on the interplay between collider searches
for MSSM Higgs bosons $(h,H,A)$, as well as some other sparticles at 
the LHC and direct detection of dark matter for the aforementioned parameter 
space. 
In particular, we discuss the possibility of determining  
both $m_{\tilde\chi^0_1} $ and $m_{\tilde\chi^0_2} $ by combining 
the data from the direct detection of dark matter experiment with that of 
the edge in the $m_{\ell^+\ell^-}$ distribution 
and the light Higgs boson mass measurement at the LHC.

We further note that 
because of $\tan\beta$ being large, copious production of $(H,A)$ 
in association with the $b$-quark would be possible at the LHC. Such an $H/A$ 
mass can be reconstructed by looking at
the $H/A \to \tau^+\tau^-$ decay mode. We also point out that 
large $\tan\beta$ and relatively low $m_{1/2}$ values 
of our
benchmark points lead to a large number of multi-b jets in association 
with missing energy signal. 
\section*{Acknowledgments}
DD thanks P2I, CNRS for the support received as a post-doctoral fellow.
DKG and MM acknowledge support from the Department of Science and Technology, India 
under grants SR/S2/HEP-12/2006 and SR/MF/PS-03/2009-VB-I respectively. DKG also thanks
Theory Division, CERN for the hospitality when part of this work was done.

\end{document}